\documentstyle[12pt]{article}
\begin{document}
\title{SO(3) Gauge model for neutrino masses and oscillations }
\author{Yue-Liang  Wu \thanks{Supported in part by Outstanding Young Scientist Research Fund of China, under 
grant No. of NSF of China: 19625514. }  \\  
\\
Institute of Theoretical Physics, Chinese Academy of Sciences, \\
 P.O. Box 2735, Beijing 100080, China } 
\date{ylwu@ITP.AC.CN}
\maketitle

\begin{abstract}
I mainly describe neutrino masses and oscillations in 
the gauge model with $SO(3)_{F}$ lepton flavor symmetry 
and with two Higgs triplets. It is shown how the maximal mixing 
between $\nu_{\mu}$ and $\nu_{\tau}$ neutrinos comes out naturally after spontaneous 
breaking of the symmetry. The nearly two-flavor mixing scenario is resulted naturally
from an approximate permutation symmetry between the two Higgs triplets.
The hierarchy between the neutrino mass-squared differences, which is needed 
for reconciling both solar and atmospheric neutrino data, leads to an almost maximal
mixing between $\nu_{e}$ and $\nu_{\mu}$ neutrinos. Thus the model favors
the intriguing bi-maximal mixing scenario.
The three Majorana neutrino masses are allowed to be nearly degenerate and large enough 
to play a significant cosmological role. The model can also lead to interesting phenomena on 
lepton-flavor violations via the $SO(3)_{F}$ gauge interactions. 
\end{abstract}

\maketitle

\newpage

\section{INTRODUCTION}

The standard model (SM) has been tested by more and more precise experiments, 
its greatest success is the gauge symmetry structure 
SU(3)$_{c}\times$ SU$_{L}(2)\times$ U$_{Y}(1)$. 
While neutrinos are assumed to be massless in the SM. Studies on neutrino 
physics have resulted in the following observations: i), 
The Super-Kamiokande data\cite{SUPERK} on atmospheric neutrino anomaly 
provide a strong evidence that neutrinos are massive; ii), The current 
Super-Kamiokande data on solar neutrino\cite{SUPERK} cannot decisively 
establish whether the deficit of the measured solar neutrino flux results from 
MSW solutions\cite{MSW} with large/small mixing angles\cite{LSM} or vacuum 
oscillation solutions\cite{VO}. iii), To describe all the neutrino phenomena such as the 
atmospheric neutrino anomaly, the solar neutrino deficit and the results 
from the LSND experiment, it is necessary to introduce a sterile 
neutrino. It indicates that with only three light neutrinos, 
one of the experimental data must be modified; iv), The current experimental 
data cannot establish whether neutrinos are Dirac-type or Majorana-type. 
The failure of detecting neutrinoless double beta decay only provides, 
for Majorana-type neutrinos, an upper bound on an `effective' 
electron neutrino mass; v), Massive neutrinos are also regarded as the best 
candidate for hot dark matter and may play an essential role in the evolution 
of the large-scale structure of the universe\cite{HDM}. 

To introduce neutrino masses and mixings, it is necessary to modify 
and go beyond the SM. As a simple extension of the standard model, 
it is of interest to introduce a flavor symmetry among the three families 
of the leptons. In the recent papers\cite{YLW1,YLW2}, we have introduced the gauged 
SO(3)$_{F}$ flavor symmetry\cite{SO3} to describe the three lepton families. Some remarkable 
features have been found to be applicable to the current interesting phenomena of 
neutrinos. After a detailed analysis on various possible scenarios, we have shown that 
the nearly degenerate neutrino mass and bi-maximal mixing scenario\cite{BMM} is the most 
favorable one in our model with two Higgs triplets\cite{YLW2} to reconcile both solar and 
atmospheric neutrino flux anomalies. In this talk, I will briefly review those interesting 
features and try to explicitly explore the naturalness for some of the features. 
To understand the naturalness of the scenario, we will pay attention to the spontaneous 
breaking of the SO(3)$_{F}$ flavor symmetry in the Higgs sector. As a consequence, the maximal
mixing between $\nu_{\mu}$ and $\nu_{\tau}$ neutrinos, which is needed for 
explaining the observed atmospheric neutrino anomaly, comes out naturally after 
spontaneous symmetry breaking. 
By considering the approximate permutation symmetry between the 
two Higgs triplets and by using the data of the neutrinoless double $\beta$ decay 
or the fact of the hierarchy between the two mass-squared differences, we then arrive at   
the nearly degenerate neutrino mass and bi-maximal mixing scenario.

\section{THE MODEL}

 For a less model-dependent analysis, we directly start from an $SO(3)_{F}\times 
SU(2)_{L}\times U(1)_{Y}$ invariant effective lagrangian with two $SO(3)_{F}$ Higgs triplets 
\begin{eqnarray}
&&{\cal L} =   \frac{1}{2}g'_{3}A_{\mu}^{k}
( \bar{L}_{i}\gamma^{\mu} (t^{k})_{ij}L_{j} 
+ \bar{e}_{R i} \gamma^{\mu}(t^{k})_{ij}e_{R j} ) \nonumber \\
&& +[ (c_{1}\varphi_{i}\varphi_{j}\chi +  c'_{1} \varphi'_{i}\varphi'_{j}\chi' + 
c''_{1} \delta_{ij}\chi'') \bar{L}_{i} \phi_{1}e_{R\ j}  \nonumber  \\
&& + [(c_{0}\varphi_{i}\varphi_{j}^{\ast} + c'_{0}\varphi'_{i}\varphi_{j}^{'\ast} 
+  c\delta_{ij} ) \bar{L}_{i} \phi_{2}\phi_{2}^{T}L_{j}^{c} + H.c. ] \nonumber \\
&&+ D_{\mu}\varphi^{\ast} D^{\mu}\varphi + D_{\mu}\varphi^{'\ast}D^{\mu}\varphi'
  - V_{\varphi} +  {\cal L}_{SM}  
\end{eqnarray}
This effective Lagrangian can be resulted from integrating out heavy particles. 
${\cal L}_{SM} $ denotes the lagrangian of the standard 
model. $\bar{L}_{i}(x) = (\bar{\nu}_{i}, \bar{e}_{i})_{L}$ 
(i=1,2,3) are the SU(2)$_{L}$ doublet leptons and $e_{R\ i}$ ($i=1,2,3$) are 
the three right-handed charged leptons. $A_{\mu}^{i}(x)t^{i}$ ($i=1,2,3$) 
are the $SO(3)_{F}$ gauge bosons with $t^{i}$ the $SO(3)_{F}$ generators and 
$g'_{3}$ is the corresponding gauge coupling constant. Here $\phi_{1}(x)$ and $\phi_{2}(x)$ 
are two Higgs doublets, $\varphi^{T}=(\varphi_{1}(x), \varphi_{2}(x), 
\varphi_{3}(x))$ and $\varphi'^{T} = (\varphi'_{1}(x), \varphi'_{2}(x), 
\varphi'_{3}(x))$ are two $SO(3)_{F}$ Higgs triplets, $\chi(x)$, 
$\chi'(x)$ and $\chi''(x)$ are three singlet scalars. The couplings $c$, $c_{a}$, $c'_{a}$ 
 ($a=0,1$) and $c''_{1}$ are dimensional constants. The structure of the above effective 
lagrangian can be obtained by imposing an additional U(1) symmetry \cite{YLW1}. 

   After the symmetry SO(3)$_{F}\times$SU(2)$_{L}\times$U(1)$_{Y}$ is broken 
down to the U(1)$_{em}$ symmetry,  mass matrices of the neutrinos 
and charged leptons get the following forms
\begin{eqnarray}
(M_{e})_{ij} & = & m_{1} \frac{\hat{\sigma}_{i}\hat{\sigma}_{j}}{\sigma^{2}} 
+ m'_{1} \frac{\hat{\sigma}'_{i}\hat{\sigma}'_{j}}{\sigma^{'2}} + m''_{1} \delta_{ij},
 \\
(M_{\nu})_{ij} & = &  m_{0} \frac{\hat{\sigma}_{i}
\hat{\sigma}_{j}^{\ast} }{2\sigma^{2}}
+  m'_{0} \frac{\hat{\sigma}'_{i}\hat{\sigma}_{j}^{'\ast} }{2\sigma^{'2}} +H.c.  
+ m_{\nu} \delta_{ij}   \nonumber
\end{eqnarray}
where the mass matrices $M_{e}$ and $M_{\nu}$ are defined in the basis
 ${\cal L}_{M} = \bar{e}_{L}M_{e}e_{R} + 
\bar{\nu}_{L}M_{\nu}\nu^{c}_{L} + H.c. $. The constants $\hat{\sigma}_{i} = 
<\varphi_{i}(x)>$ and $\hat{\sigma}'_{i} = <\varphi'_{i}(x)>$ 
are the vacuum expectation values (VEVs) of the two Higgs triplets with 
$\sigma^{2} = \sum_{i=1}^{3}|\hat{\sigma}_{i}|^{2}$ and $\sigma^{'2} = 
\sum_{i=1}^{3}|\hat{\sigma}'_{i}|^{2}$. Here $m_{1}$, $m'_{1}$ and $m''_{1}$ 
 as well as $m_{\nu}$, $m_{0}$ and $m'_{0}$ are mass parameters. 

  For simplicity, we only present here the Higgs potential for the SO(3)$_{F}$ Higgs triplets  
\begin{eqnarray}
V_{\varphi} & = & \frac{1}{2} \mu^{2} (\varphi^{\dagger}\varphi ) + 
\frac{1}{2} \mu^{'2} (\varphi'^{\dagger}\varphi' )
 +\frac{1}{4}\lambda (\varphi^{\dagger}\varphi)^{2}  \nonumber \\
& + & \frac{1}{4}\lambda' (\varphi'^{\dagger}\varphi')^{2} 
 +  \frac{1}{2}\kappa_{1} (\varphi^{\dagger}\varphi )( \varphi'^{\dagger}\varphi') \nonumber \\ 
& + & \frac{1}{2} \kappa_{2} (\varphi^{\dagger}\varphi')( \varphi'^{\dagger}\varphi ) .
\end{eqnarray}
where we have omitted terms involving other Higgs fields since those terms 
will not change our conclusions. 
  
   In our considerations, the SO(3)$_{F}$ flavor symmetry is treated to be a gauge symmetry, 
thus the complex SO(3)$_{F}$ Higgs triplet fields $\varphi_{i}(x)$ ($\varphi'_{i}(x)$) can always 
be expressed in terms of the three rotational fields $\eta_{i}(x)$ ($\eta'_{i}(x)$)
and three amplitude fields $\rho_{i}(x)$ ($\rho'_{i}(x)$) ($i=1,2,3$), i.e., 
$\varphi(x) = O(x)\rho(x)$ and $\varphi'(x) = O'(x)\rho'(x)$
with  $O(x)\equiv e^{i\eta_{i}(x)t^{i}}$ and $O'(x)\equiv e^{i\eta'_{i}(x)t^{i}}$ 
$\in $ $SO(3)_{F}$ being the $SO(3)_{F}$ rotational fields.
This is analogous to SU(2) gauge symmetry, the complex SU(2) doublet 
scalar field can always be expressed in terms of three SU(2) `rotational'  fields and one 
amplitude field. As the SO(3) rotation matrix is real, which is unlike the SU(2) rotation 
matrix that is complex, one of the three amplitude fields of the complex SO(3) triplet  
scalar  must be a pure imaginary field so that one can generate the complex SO(3) triplet scalar 
fields $\varphi_{i}(x)$ by the SO(3) field  $O(x)= e^{i\eta_{i}(x)t^{i}} \in $ SO(3) action on 
the three amplitude fields. Explicitly, one has 
\begin{equation} 
 \left( \begin{array}{c}
  \varphi_{1}(x) \\
  \varphi_{2}(x)   \\
  \varphi_{3}(x)   \\
\end{array} \right) = e^{i\eta_{i}(x)t^{i}} \frac{1}{\sqrt{2}}
\left( \begin{array}{c}
  \rho_{1}(x) \\
  i\rho_{2}(x)   \\
  \rho_{3}(x)   \\
\end{array} \right) 
\end{equation}
Similar form is for $\varphi'(x)$. SO(3)$_{F}$ gauge symmetry allows one to 
remove three degrees of freedom from six rotational fields. 
Making $SO(3)_{F}$ gauge transformations: $(\varphi(x), \varphi'(x) )
 \rightarrow O^{T}(x) (\varphi (x), \varphi'(x))$, and assuming that
only the amplitude fields get VEVs after spontaneous breaking of the $SO(3)_{F}$ 
flavor symmetry, namely  $<\rho_{i}(x)> = \sigma_{i}$ and $<\rho'_{i}(x)> = \sigma'_{i}$, 
we then obtain the following equations from minimizing the Higgs potential $V_{\varphi}$ 
\begin{eqnarray}
& & \omega^{2} \sigma_{i} 
+ \kappa_{2} \sum_{j=1}^{3}(\sigma_{j}\sigma'_{j}) \sigma'_{i}   = 0 , \nonumber \\
& & \omega^{'2} \sigma'_{i}  
+ \kappa_{2} \sum_{j=1}^{3}(\sigma_{j}\sigma'_{j}) \sigma_{i} = 0  
\end{eqnarray} 
with $\omega^{2} = \mu^{2} + \lambda \sigma^{2} + \kappa_{1} \sigma^{'2}$ and 
$\omega^{'2} = \mu^{'2} + \lambda' \sigma^{'2} + \kappa_{1} \sigma^{2} $.
To find out possible constraints, it is useful to set $\sigma'_{i} = \xi_{i} \sigma_{i}$ 
for $\sigma_{i}\neq 0$ with $i=1,2,3$ and $\sigma^{'2} = \xi \sigma^{2}$.
When $\xi_{1} = \xi_{2} = \xi_{3}=\sqrt{\xi} $, the two $SO(3)_{F}$ 
Higgs triplets $\varphi(x)$ and $\varphi'(x)$ are parallel in the model and 
the introduction of the second Higgs triplet becomes trivial.
For the general and nontrivial case, it is easy to check that when   
$\xi_{i}=\xi_{j} \neq \xi_{k}\equiv \xi_{i} - \xi_{0}$ with $i\neq j \neq k$,  
one arrives at the strong constraints from the minimum conditions in eq.(5)
\begin{eqnarray}
& & \sum_{i=1}^{3} \sigma_{i}\sigma'_{i} = \sum_{i=1}^{3}\xi_{i} \sigma_{i}^{2} = 0 , \nonumber \\
& & \omega^{2}= \mu^{2} + \lambda \sigma^{2} + \kappa_{1} \sigma^{'2} = 0, \\
& & \omega^{'2} = \mu^{'2} + \lambda' \sigma^{'2} + \kappa_{1} \sigma^{2}. \nonumber 
\end{eqnarray} 
For convenience of discussions, we make, without lossing generality, the
convention that $\xi_{1}=\xi_{2} \neq \xi_{3}\equiv \xi_{1} - \xi_{0}$. 
Thus from the above constraints, we obtain the solutions  
\begin{equation}
\xi = \xi_{1}(\xi_{0}-\xi_{1}), \  \  \xi - \xi_{1}^{2}\tan^{2}\theta_{2}  =0
\end{equation}
with $\tan^{2}\theta_{2}=\sigma_{12}^{2}/\sigma_{3}^{2}$ and 
$\sigma_{12}^{2}= \sigma_{1}^{2} + \sigma_{2}^{2}$.  
   Furthermore, one must check the minimum conditions directly from 
the Higgs potential at the minimizing point. It is easy to see that 
\begin{eqnarray}
V_{\varphi}|_{min} & = & -\sigma^{4}(\lambda + \lambda'\xi^{2} + 2\kappa_{1}\xi )/4
\end{eqnarray}
This implies that to have a minimum potential energy $V_{\varphi}|_{min}$ for 
varying $\xi$, the value of $\xi$ is required to be maximal 
for positive coupling constants $\lambda$'s and $\kappa_{1}$. From such a requirement, it is seen 
that for the given $\xi_{0}$ in eq.(7), the maximum condition for 
$\xi$ lead to the solution $\xi_{1}= \xi_{0}/2 = \sqrt{\xi} = \xi_{2} = -\xi_{3}$, namely
\begin{eqnarray}
& & \sigma'_{1}= \sqrt{\xi} \sigma_{1}, \    \sigma'_{2}= \sqrt{\xi} \sigma_{2}, 
\   \sigma'_{3}= -\sqrt{\xi} \sigma_{3}, \nonumber \\ 
& &  \sigma_{3}^{2} = \sigma_{1}^{2} + \sigma_{2}^{2} 
\  or  \  \theta_{2} = \pi/4  
\end{eqnarray}
where $\xi$ is given as a function of the coupling constants in the Higgs potential, 
$\xi = (-\lambda\mu^{'2} + \kappa_{1}\mu^{2})/(-\lambda'\mu^{2} + \kappa_{1}\mu^{'2})$. 
Thus the VEVs are completely determined by the Higgs potential.

\section{NEUTRINO MASSES AND OSCILLATIONS}

  It is interesting to note that with the above relations the mass matrices of the neutrinos and 
charged leptons are greatly simplified to the following nice forms
\begin{eqnarray}
M_{e} & = & \frac{m_{1}}{2}\left( \begin{array}{ccc}
  s_{1}^{2} & ic_{1}s_{1} & s_{1}  \\
   ic_{1}s_{1} & -c_{1}^{2} &  ic_{1} \\
  s_{1} & ic_{1} & 1 \\ 
\end{array} \right) \nonumber \\
 & + &  \frac{m'_{1}}{2}\left( \begin{array}{ccc}
 s_{1}^{2} & ic_{1}s_{1} & -s_{1}  \\
   ic_{1}s_{1} & -c_{1}^{2} &  -ic_{1} \\
  -s_{1} & -ic_{1} & 1 \\ 
\end{array} \right) \\
& + & \frac{m''_{1}}{2} \left( \begin{array}{ccc}
  1 & 0 & 0  \\
  0 & 1  & 0 \\
 0 & 0 &  1 \\ 
\end{array} \right) \nonumber 
\end{eqnarray}
and
\begin{equation}
M_{\nu} = m_{\nu}\left( \begin{array}{ccc}
  1 + \delta_{+}s_{1}^{2}  & 0 & \delta_{-}s_{1}  \\
  0 & 1 + \delta_{+}c_{1}^{2}  &  0 \\
\delta_{-}s_{1} & 0
 & 1 + \delta_{+}  \\ 
\end{array} \right)
\end{equation}
with $\delta_{\pm} = (m_{0}\pm m'_{0})/2m_{\nu}$ . 

  It is remarkable that the two nondiagonal mass matrices in $M_{e}$ can be diagonalized by a unitary 
bi-maximal mixing matrix $U_{e}$ via $M'_{e} = U_{e}^{\dagger} M_{e} U_{e}^{\ast}$. Here
\begin{equation} 
M'_{e}  = \left( \begin{array}{ccc}
  0 & 0 & 0  \\
  0 & m'_{1}& 0  \\
  0 & 0 & m_{1}   \\
\end{array} \right) + m''_{1}U_{e}^{\dagger}U_{e}^{\ast}
\end{equation}
and 
\begin{equation}
U_{e}^{\dagger}=\left( \begin{array}{ccc}
  ic_{1} & -s_{1} & 0  \\
 c_{2}s_{1} & -i\frac{1}{\sqrt{2}} c_{1} & - \frac{1}{\sqrt{2}} \\
 \frac{1}{\sqrt{2}} s_{1} & -i \frac{1}{\sqrt{2}} c_{1} & \frac{1}{\sqrt{2}}  \\
\end{array} \right)
\end{equation}
where $U_{e}^{\dagger}U_{e}^{\ast}$ has the following explicit form 
\begin{eqnarray}
U_{e}^{\dagger}U_{e}^{\ast} & = & \left( \begin{array}{ccc}
  0 & \sqrt{2} ic_{1}s_{1} & \sqrt{2} ic_{1}s_{1}  \\
   \sqrt{2}ic_{1}s_{1} & \frac{1}{2} & - \frac{1}{2}  \\
  \sqrt{2}ic_{1}s_{1} & - \frac{1}{2}
 & \frac{1}{2}  \\ 
\end{array} \right) \nonumber \\
& + & (s_{1}^{2}-c_{1}^{2})\left( \begin{array}{ccc}
  1 & 0 & 0  \\
  0 & \frac{1}{2} & \frac{1}{2}  \\
  0 & \frac{1}{2} & \frac{1}{2}  \\
\end{array} \right)
\end{eqnarray}
The hierarchical structure of the charged lepton mass implies that 
$m''_{1} << m'_{1} << m_{1}$, it is then not difficult to see that 
the matrix $M'_{e}$ will be further diagonalized by a unitary matrix 
$U'_{e}$ via $D_{e} = U_{e}^{'\dagger} M'_{e} U_{e}^{'\ast} =  
U_{e}^{'\dagger} U_{e}^{\dagger} M_{e} U_{e}^{\ast} U_{e}^{'\ast}$ with 
\begin{equation} 
D_{e}  = \left( \begin{array}{ccc}
  m_{e} & 0 & 0  \\
  0 & m_{\mu}& 0  \\
  0 & 0 & m_{\tau}   \\
\end{array} \right) 
\end{equation}
and 
\begin{equation}
U_{e}^{'\dagger}=\left( \begin{array}{ccc}
  1 + \epsilon_{2} & i\epsilon_{2} & i\epsilon_{1}  \\
 i\epsilon_{2} & 1 + \epsilon_{2} & \epsilon_{1} \\
 i\epsilon_{1} & \epsilon_{1} & 1 + \epsilon_{1} \\
\end{array} \right)
\end{equation}
with  $\epsilon_{1}= O(\sqrt{m_{e}m_{\mu}}/m_{\tau}) $ and 
$\epsilon_{2}= O(\sqrt{m_{e}/m_{\mu}})$. 
Where $m_{e}\simeq m_{1}^{''2}/m'_{1} $, $m_{\mu}= m'_{1} + O(m''_{1}) $ and 
$m_{\tau}= m_{1} + O(m''_{1})$ define the three charged lepton masses. 
This indicates that the unitary matrix $U'_{e}$ does not significantly 
differ from the unit matrix.
 
 The neutrino mass matrix can be easily diagonalized 
by an orthogonal matrix $O_{\nu}$ via $O_{\nu}^{T}M_{\nu}O_{\nu}$ 
\begin{equation}
O_{\nu}=\left( \begin{array}{ccc}
  c_{\nu} & 0 & s_{\nu}  \\
 0 & 1 & 0 \\
 - s_{\nu} & 0 & c_{\nu}  \\
\end{array} \right)
\end{equation}
with $s_{\nu}\equiv \sin\theta_{\nu}$ and 
$\tan2\theta_{\nu} = 2\delta_{-}s_{1}/(\delta_{+}c_{1}^{2})$

When going to the physical mass basis of the neutrinos and charged leptons, 
we then obtain the CKM-type lepton mixing matrix $U_{LEP}$  
appearing in the interactions of the charged weak gauge bosons and leptons, i.e.,
${\cal L}_{W} = \bar{e}_{L}\gamma^{\mu}U_{LEP} \nu_{L} W_{\mu}^{-} + H.c. $. 
Explicitly, we have $U_{LEP} = U_{e}^{'\dagger}U_{e}^{\dagger}O_{\nu}$ with 
\begin{eqnarray}
U_{LEP} & = & U_{e}^{'\dagger}\left( \begin{array}{ccc}
  ic_{1}c_{\nu} & -s_{1} & 0  \\
 \frac{1}{\sqrt{2}}s_{1}c_{\nu} & -i\frac{1}{\sqrt{2}}c_{1} & 
-\frac{1}{\sqrt{2}}c_{\nu} \\
 \frac{1}{\sqrt{2}}s_{1}c_{\nu} & -i\frac{1}{\sqrt{2}}c_{1} & 
\frac{1}{\sqrt{2}}c_{\nu}  \\
\end{array} \right) \nonumber \\
& + & U_{e}^{'\dagger} \left( \begin{array}{ccc}
  0 & 0 & ic_{1}  \\
 \frac{1}{\sqrt{2}} & 0 & \frac{1}{\sqrt{2}}s_{1} \\
 - \frac{1}{\sqrt{2}} & 0 & \frac{1}{\sqrt{2}}s_{1}  \\
\end{array} \right) s_{\nu} 
\end{eqnarray}
The three neutrino masses are found to be
\begin{eqnarray}
m_{\nu_{e}} & = &  m_{\nu}[ 1 + \delta_{+}s_{1}^{2} 
- \delta_{+}c_{1}^{2} t_{\nu}^{2}/(1-t_{\nu}^{2}) \  ] 
\nonumber \\
m_{\nu_{\mu}} & = &  m_{\nu}[ 1 + \delta_{+}c_{1}^{2} ]   \\
m_{\nu_{\tau}} & = &  m_{\nu}[ 1 + \delta_{+}  
+ \delta_{+}c_{1}^{2} t_{\nu}^{2}/(1-t_{\nu}^{2}) \    ] \nonumber 
\end{eqnarray}
with $t_{\nu} \equiv s_{\nu}/c_{\nu}$. 

  The similarity between the Higgs triplets $\varphi(x)$ and $\varphi'(x)$
naturally motivates us to consider an approximate (and softly broken) permutation 
symmetry between them. This implies that $m_{0}\simeq m'_{0}$. As a consequence, 
one has $|\delta_{-}|<< 1$. To a good approximation, the mass-squared 
differences are given by  
\begin{eqnarray}
& & \Delta m_{\mu e}^{2} \simeq  2 m_{\nu}^{2} \delta_{+} 
(c_{1}^{2}-s_{1}^{2} + (\delta_{-}s_{1}/\delta_{+}c_{1})^{2} ), \nonumber \\
& & \Delta m_{\tau\mu}^{2} \simeq 
2 m_{\nu}^{2} \delta_{+} s_{1}^{2}
\end{eqnarray}
with $\Delta m_{ij}^{2} \equiv m_{\nu_{i}}^{2} - m_{\nu_{j}}^{2}$. 

On the other hand, from the fact that the failure of detecting neutrinoless 
double beta decay provide 
bounds on an effective electron neutrino mass $<m_{\nu_{e}}> = \sum_{i} 
m_{\nu_{i}}(U_{LEP})_{ei}^{2} < 0.2$ eV\cite{DB}. We then obtain
\begin{equation}
 <m_{\nu_{e}}> \simeq m_{\nu}|s_{1}^{2} - c_{1}^{2}| < 0.2 eV
\end{equation}
Assuming that neutrino masses are large enough to play an essential 
role in the evolution of the large-scale structure of the universe, we may set 
$m_{\nu} \sim 1$ eV, thus the above constraint will result in the following bound on the 
mixing angle $\theta_{1}$ 
\begin{equation}
|s_{1}^{2}- c_{1}^{2}| < 0.2
\end{equation}
which implies that 
$\Delta m_{\mu e}^{2}/\Delta m_{\tau\mu}^{2} < 0.4$. To explain the 
solar neutrino data, the allowed range of the ratio is 
$\Delta m_{\mu e}^{2}/\Delta m_{\tau\mu}^{2}
 \sim 10^{-2}-10^{-8}$. Here the large value is for matter-enhanced MSW solution\cite{MSW} 
with large mixing angle\cite{LSM} and the small value for the vacuum 
oscillation solutions\cite{VO}. As a consequence, the neutrino mixing between 
$\nu_{e}$ and $\nu_{\mu}$ becomes almost maximal
\begin{equation} 
\sin^{2}2\theta_{1} > 0.998
\end{equation} 

  With the hierarchical feature in $\Delta m^{2}$,  
formulae for the oscillation probabilities in vacuum are greatly simplified to be
\begin{eqnarray}
& & P_{\nu_{e}\rightarrow \nu_{e}}|_{solar} \simeq  1 - 
\sin^{2}(\frac{\Delta m_{\mu e}^{2}L}{4E}) \nonumber \\
& & P_{\nu_{\mu}\rightarrow \nu_{\mu}}|_{atm.} \simeq 1 -
\sin^{2}(\frac{\Delta m_{\tau\mu}^{2}L}{4E}) \\
& & P_{\nu_{\beta}\rightarrow \nu_{\alpha}} \simeq 4|U_{\beta 3}|^{2}|U_{\alpha 3}|^{2}
\sin^{2}(\frac{\Delta m_{\tau\mu}^{2}L}{4E}) \nonumber \\
& & P_{\nu_{\mu}\rightarrow \nu_{e}}/P_{\nu_{\mu}\rightarrow \nu_{\tau}}|_{atm.}
\simeq (\Delta m_{\mu e}^{2}/\Delta m_{\tau\mu}^{2}) << 1 \   . \nonumber 
\end{eqnarray}
This may present the simplest scheme for reconciling both 
solar and atmospheric neutrino fluxes via oscillations of three neutrinos. But 
it needs a strong fine-tuning. 

  When going back to the weak gauge and charged-lepton mass basis, 
the neutrino mass matrix is given by $M_{\nu} = U_{e}^{' \dagger} \hat{M}_{\nu} 
U_{e}^{' \ast}$. Where $\hat{M}_{\nu}$ has the following general and interesting form 
\begin{eqnarray}
&\hat{M}_{\nu} & \simeq \hat{m}_{\nu} \left( \begin{array}{ccc}
  0 & \frac{1}{\sqrt{2}}i & \frac{1}{\sqrt{2}}i  \\
   \frac{1}{\sqrt{2}}i & \frac{1}{2} &  - \frac{1}{2}  \\
  \frac{1}{\sqrt{2}}i &  - \frac{1}{2} &  \frac{1}{2}  \\ 
\end{array} \right) \nonumber \\
& + &  \frac{\hat{m}_{\nu}\hat{\delta}_{-}}{2}\left( \begin{array}{ccc}
 0 & -\frac{1}{\sqrt{2}}i & \frac{1}{\sqrt{2}}i  \\
   -\frac{1}{\sqrt{2}}i & -1 &  0  \\
  \frac{1}{\sqrt{2}}i & 0 &  1 \\  
\end{array} \right)  \\
& + & \frac{\hat{m}_{\nu}\hat{\delta}_{+}}{2}\left( \begin{array}{ccc}
 0 & 0 & 0  \\
   0 & 1 &  -1  \\
  0 & -1 &  1 \\  
\end{array} \right) + (s_{1}^{2} - c_{1}^{2}) M'_{\nu}. \nonumber 
\end{eqnarray}
with $\hat{m}_{\nu}= m_{\nu}(1 + \delta_{+}c_{1}^{2})$, 
$\hat{\delta}_{-}= \delta_{-}s_{1}/(1 + \delta_{+}c_{1}^{2})$ and  
$\hat{\delta}_{+}= \delta_{+}s_{1}^{2}/(1 + \delta_{+}c_{1}^{2})$.
For $ |s_{1}^{2} - c_{1}^{2}| \sim 10^{-2}$, the neutrino masses can be approximately 
degenerate and large enough  ( $\hat{m}_{\nu}= O(1)$ eV) to play a 
significant cosmological role.

\section{SO(3) GAUGE INTERACTIONS AND LEPTON-FLAVOR VIOLATIONS}

We now come to discuss SO(3) gauge interactions.  After the SO(3) gauge symmetry is 
spontaneously broken down, the gauge fields $A_{\mu}^{i}$ receive masses by 
`eating' three of the rotational fields. For the SO(3) vacuum structure given 
above, $A_{\mu}^{1}$ and $A_{\mu}^{3}$ are not in the mass eigenstates since 
they mix each other.

 The mass matrix of gauge fields $A_{\mu}^{i}$ is found to be
\[
 M_{F}^{2} = m_{F}^{2}\left( \begin{array}{ccc}
  1  & 0  & -\frac{s_{1}\xi_{-}}{\xi_{+}} \\ 
    0 & 1+c_{1}^{2}+ \alpha & 0  \\ 
 -\frac{s_{1}\xi_{-}}{\xi_{+}} & 0  &  1+s_{1}^{2}-\alpha  
\end{array} \right) 
\]
with $ m_{F}^{2}= \xi_{+}g^{'2}_{3}\sigma^{2}/8$, $\xi_{\pm}=(1 \pm \xi)/2$ and 
$\alpha = (s_{1}^{2}-c_{1}^{2})/2\xi_{+}$. 
This mass matrix is diagonalized by an orthogonal matrix $O_{F}$ 
via $O_{F}^{T}M_{F}^{2}O_{F}$. Denoting the physical gauge fields as $F_{\mu}^{i}$, 
we then have $A_{\mu}^{i} = O_{F}^{ij}F_{\mu}^{j}$, i.e., 
\begin{equation} 
\left( \begin{array}{c} A_{\mu}^{1} \\ A_{\mu}^{2} \\ A_{\mu}^{3} \\ \end{array} 
\right)  = \left( \begin{array}{ccc} c_{F} & 0  & -s_{F} \\ 0 & 1 & 0  \\ s_{F}  
&  0 & c_{F}  \\ \end{array} \right) \left( \begin{array}{c} F_{\mu}^{1} \\ 
F_{\mu}^{2} \\ F_{\mu}^{3} \\ \end{array} \right) 
\end{equation} 
with $s_{F}\equiv \sin \theta_{F}$ and 
\begin{equation}
\tan 2\theta_{F}  =  \frac{4s_{1}\xi_{-}}{2\xi_{+}s_{1}^{2} + c_{1}^{2}-s_{1}^{2}}
\end{equation}
Masses of the three physical gauge bosons $F_{\mu}^{i}$ are found to be 
\begin{eqnarray}
m_{F_{1}}^{2} & = & m_{F}^{2} [ (2+s_{1}^{2})\xi_{+} + \frac{c_{1}^{2}-s_{1}^{2}}{2} \nonumber \\
& - & (\xi_{+}s_{1}^{2} + \frac{c_{1}^{2}-s_{1}^{2}}{2} \sqrt{1 + \tan^{2}2\theta_{F}}\ ], \nonumber \\
m_{F_{2}}^{2} & = &   m_{F}^{2} [ 2(1+c_{1}^{2})\xi_{+} + s_{1}^{2}-c_{1}^{2}\ ], \\
m_{F_{3}}^{2} & = &  m_{F}^{2} [ (2+s_{1}^{2})\xi_{+} + \frac{c_{1}^{2}-s_{1}^{2}}{2} \nonumber \\
& + & (\xi_{+}s_{1}^{2} + \frac{c_{1}^{2}-s_{1}^{2}}{2} \sqrt{1 + \tan^{2}2\theta_{F}}\ ]. \nonumber
\end{eqnarray}
  In the physical mass basis of the leptons and gauge bosons, 
the gauge interactions of the leptons are given by the following form 
\begin{eqnarray}
{\cal L}_{F} & = &  \frac{1}{2}g'_{3} F_{\mu}^{i}\bar{\nu}_{L}t^{j}
O_{F}^{ji}\gamma^{\mu}\nu_{L} \nonumber \\
& + &  \frac{1}{2}g'_{3} F_{\mu}^{i}
\left( \bar{e}_{L} V_{e}^{i}\gamma^{\mu}e_{L} - \bar{e}_{R} V_{e}^{i \ast}
\gamma^{\mu}e_{R}\right) 
\end{eqnarray}
with $V_{e}^{i} = U^{'\dagger}\hat{V}_{e}^{i}U'_{e}$.
 Here $\hat{V}_{e}^{i}=  K_{e}^{j}O_{F}^{ji}$, i.e., 
\begin{eqnarray}
\hat{V}_{e}^{1} & = & \cos\theta_{F}K_{e}^{1} + \sin\theta_{F}K_{e}^{3}, \nonumber \\
 \hat{V}_{e}^{2} & = & K_{e}^{2}, \\
\hat{V}_{e}^{3} & = & -\sin\theta_{F}K_{e}^{1} + \cos\theta_{F}K_{e}^{3} \nonumber 
\end{eqnarray} 
where
\begin{eqnarray}
K_{e}^{1} & = & c_{1}s_{1}\left( \begin{array}{ccc}
  2 & 0 & 0 \\
   0 & 1 &  1 \\
  0 &  1  & 1  \\ 
\end{array} \right) \nonumber \\
& + & i\frac{1}{\sqrt{2}}(s_{1}^{2}-c_{1}^{2})
\left( \begin{array}{ccc}
 0 & 1 & 1  \\
   -1 & 0 &  0  \\
  -1 & 0 &  0 \\  
\end{array} \right), \nonumber \\ 
K_{e}^{2} & = & \left( \begin{array}{ccc}
  0 & \frac{1}{\sqrt{2}}c_{1} & -\frac{1}{\sqrt{2}}c_{1}  \\
   \frac{1}{\sqrt{2}}c_{1} & 0 &  is_{1}  \\
 -\frac{1}{\sqrt{2}}c_{1} &  -is_{1} & 0 \\ 
\end{array} \right), \\
K_{e}^{3} & = & \left( \begin{array}{ccc}
  0 & i\frac{1}{\sqrt{2}}s_{1} & -i\frac{1}{\sqrt{2}}s_{1}  \\
   -i\frac{1}{\sqrt{2}}s_{1}  & c_{1} &  0  \\
 i\frac{1}{\sqrt{2}}s_{1} &   & -c_{1}  \\ 
\end{array} \right). \nonumber
\end{eqnarray} 
Thus the $SO(3)_{F}$ gauge interactions allow lepton flavor violating processes. For 
$\mu \rightarrow 3e$ decay, its branch ratio is found
\begin{equation}
Br(\mu \rightarrow 3e) = \left(\frac{v}{\sigma}\right)^{4}
 \frac{2\xi_{-}^{2}}{(3\xi_{+}^{2} - \xi_{-}^{2} )^{2} } 
\end{equation}
with $v=246$GeV. For $\sigma \sim 10^{3} v$, the branch ratio could be very close to 
the present experimental upper bound $Br(\mu \rightarrow 3e) < 1 \times 10^{-12}$ \cite{LFV}.
Thus when taking the mixing angle $\theta_{F}$ and the coupling constant $g'_{3}$ for the 
$SO(3)_{F}$ gauge bosons to be at the same order of magnitude as those for the 
electroweak gauge bosons, we find that masses of the $SO(3)_{F}$ gauge bosons 
are at the order of magnitudes $m_{F_{i}} \sim 10^{3} m_{W}\simeq 80 $ TeV.

\section{CONCLUSIONS}

Based on the gauge model with $SO(3)_{F}$ lepton flavor symmetry 
and two Higgs triplets, we have shown how the maximal mixing 
between $\nu_{\mu}$ and $\nu_{\tau}$ neutrinos comes out naturally after spontaneous 
breaking of the symmetry. We have also shown that a two-flavor mixing scenario 
can be naturally resulted from an approximate permutation symmetry
between the two Higgs triplets. An almost maximal mixing between $\nu_{e}$ and 
$\nu_{\mu}$ neutrinos has been found to be a natural conseqnece of the hierarchical feature 
between the neutrino mass-squared differences. Thus the model favors
the almost bi-maximal mixing scenario\cite{YLW}. The model allows
three Majorana neutrino masses to be nearly degenerate and large enough 
to play a significant cosmological role. The $SO(3)_{F}$ gauge interactions may lead to 
interesting phenomena on lepton-flavor violations.

%\end{references}


\begin{thebibliography}{99}
%\begin{references}
%\bibitem[*]{byline}  
\bibitem{SUPERK} Recent experimental results from 
Super-Kamiokande Collaboration, Y.Fukuda {\it et al.}, Phys. Rev. Lett. 
{\bf 82} (1999) 1810; Y.Fukuda {\it et al.}, Phys. Rev. Lett. {\bf 82} (1999) 2430;
 T. Kajita, in this Proceedings. 
\bibitem{MSW} L. Wolfenstein, Phys. Rev. {\bf D17} (1978) 2369;   
 S.P. Mikheyev and A. Yu. Smirnov, Yad.Fiz. {\bf 42} (1985) 1441 
[Sov. J. Nucl. Phys. {42} (1985) 913].
\bibitem{LSM} J.N. Bahcall, P. Krastev and A.Yu. Smirnov, hep-ph/9905220; 
K. Inoue, Talk at 8th Int. Workshop ``Neutrino Telescopes'', Venice, February, 1999; 
J.N. Bahcall and P.I. Krastev, Phys. Lett. {\bf B436} (1998) 243. 
\bibitem{VO}   V. Barger and K. Whisnant, hep-ph/9903262; 
J.N. Bahcall, P. Krastev and A.Yu. Smirnov, Phys. Rev. {\bf D58} (1998) 096016;
 V. Berezinsky, G. Fiorentini and M. Lisia, hep-ph/9811352.
\bibitem{HDM}J.R. Primack and M. A. K. Gross, astro-ph/9810204; 
 J.R. Primack, J. Holtzman, A. Klypin and D.O. Caldwell, 
Phys. Rev. Lett. {\bf 74} (1995) 2160.
\bibitem{YLW1} Y.L. Wu, hep-ph/9810491, to appear in Phys. Rev. D..
\bibitem{YLW2} Y.L. Wu, hep-ph/9901245, to appear in Eur. Phys. J. C.; 
hep-ph/9901320; hep-ph/9905222.
\bibitem{SO3} Literatures concerning SO(3) flavor symmetry see:
 D.O. Caldwell and R.N. Mohapatra, Phys. Rev. {\bf D40} (1994) 3477;
P. Bamert and C.P. Burgers, Phys. Lett. {\bf B329} (1994) 289; 
A.S. Joshipura, Z. Phys. {\bf C64} (1994) 31; 
C. Carone and M. Sher, Phys. Lett. {\bf B420} (1998) 83;
 E. Ma, hep-ph/9812344; C. Wetterich, hep-ph/9812426;  
R. Barbieri, L.J. Hall, G.L. Kane and G.G. Ross, hep-ph/9901228; 
A. Ghosal, hep-ph/9905470.   
\bibitem{BMM}  See for example: 
V. Barger, S. Pakavasa, T. Weiler and K. Whisnant, 
Phys. Lett. {\bf B437} (1998) 107. 
\bibitem{DB} The current experimental bound is $(M_{\nu})_{ee} < 0.2$ eV, see:  
L. Baudis et al., hep-ex/9902014.
\bibitem{LFV}  See summary talk by D. Bryman at ICHEP98, Vancouver, Canada, 1998.
\bibitem{YLW} Most recently, we have shown that the intriguing nearly degenerate neutrino mass 
and bi-maximal mixing scenario can naturally be derived from the SO(3)$_{F}$ gauge model
with three Higgs triplets; hep-ph/9906435.
\end{thebibliography}
\end{document}